\documentclass[reprint,
amsmath,amssymb,11pt,reprint,aps,prl,showpacs,floatfix]{revtex4-2}

\usepackage{lineno}
\usepackage[english]{babel}
\usepackage{multirow}
\usepackage[thinspace]{SIunits}
\usepackage{xspace}
\usepackage{ulem}
\usepackage{soul}
\usepackage{graphicx}
\graphicspath{{figures//}} 
\usepackage{dcolumn}
\usepackage{bm}
\usepackage{color}
\usepackage{wrapfig}
\usepackage{hyperref}
\usepackage{array}
\usepackage{booktabs}
\usepackage{multirow}
\usepackage{amssymb}

\usepackage{marvosym}
\hypersetup{
    unicode=false,          
    pdftoolbar=true,        
    pdfmenubar=true,        
    pdffitwindow=false,     
    pdfstartview={FitH},    
    pdftitle={My title},    
    pdfauthor={Author},     
    pdfsubject={Subject},   
    pdfcreator={Creator},   
    pdfproducer={Producer}, 
    pdfkeywords={keyword1} {key2} {key3}, 
    pdfnewwindow=true,      
    colorlinks=true,       
    linkcolor=blue,          
    citecolor=blue,        
    filecolor=blue,      
    urlcolor=blue,       
    plainpages=false        
}

\newcommand{\RNS}{\mbox{Rb$_{2}$Ni$_{3}$S$_{4}$}\xspace}

\newcommand{\Agl}{$A_{g}^{(1)}$\xspace}
\newcommand{\Agz}{$A_{g}^{(2)}$\xspace}

\newcommand{\Blg}{$B_{1g}$\xspace}

\newcommand{\wn}{\rm{cm}$^{-1}$\,}

\begin{document}

\title{ \Large Emergent Vibronic Spectral Hierarchy in a Kagome Flat-Band Insulator}

\author{Jun Shu\textsuperscript{1,2,*}, 
Jun Shen\textsuperscript{1,\Letter}, 
Yanmin Zhang\textsuperscript{3,*},
Hong Du\textsuperscript{4,*},
Qingsong Wang\textsuperscript{1,5},
Zeyuan Wang\textsuperscript{4},
Bin Wang\textsuperscript{6},
Zeliang Xu\textsuperscript{2,7},
Dengjing Wang\textsuperscript{2},
Hengfu Lin\textsuperscript{2,7},
Zunming Lu\textsuperscript{5},
Lei Qin\textsuperscript{3},
Jie Yuan\textsuperscript{8,9},
Jinbo Peng\textsuperscript{4,\Letter},
Zhida Song\textsuperscript{6},
Fedor V Kusmartsev\textsuperscript{10},
Anna Kusmartseva\textsuperscript{11},
Kui Jin\textsuperscript{8,9,12},
Ruidan Zhong\textsuperscript{4,\Letter},
and Ge He\textsuperscript{1,13,\Letter}
}

\affiliation{
\textsuperscript{1} School of Mechanical Engineering, Beijing Institute of Technology, Beijing 100081, China\\
\textsuperscript{2} Department of Applied Physics, Wuhan University of Science and Technology, Wuhan 430081, China\\
\textsuperscript{3} Beijing Key Laboratory for Sensor, Beijing Information Science and Technology University, Beijing 100192, China\\
\textsuperscript{4} Tsung-Dao Lee Institute, School of Physics and Astronomy, Shanghai Jiao Tong University, Shanghai 200240, China\\
\textsuperscript{5} School of Materials Science and Engineering, Hebei University of Technology, Tianjin 300130, China\\
\textsuperscript{6} International Center for Quantum Materials, School of Physics, Peking University, Beijing 100871, China\\
\textsuperscript{7} Hubei Province Key Laboratory of System Science in Metallurgical Process, Wuhan 430065, China\\
\textsuperscript{8} Beijing National Laboratory for Condensed Matter Physics, Institute of Physics, Chinese Academy of Sciences, Beijing 100190, China\\
\textsuperscript{9} School of Physical Sciences, University of Chinese Academy of Sciences, Beijing 100049, China\\
\textsuperscript{10} College of Engineering and Physical Sciences, Khalifa University, PO Box, 51133, Abu Dhabi, United Arab Emirates\\
\textsuperscript{11} Physics Department, Loughborough University, Loughborough LE11 3TU, UK\\
\textsuperscript{12} Songshan Lake Materials Laboratory, Dongguan, Guangdong 523808, China\\
\textsuperscript{13} Beijing Key Laboratory of Quantum Matter State Control and Ultra-Precision Measurement Technology\mbox{,} Beijing Institute of Technology\mbox{,} Beijing 100081\mbox{,} China\\
\textsuperscript{*} These authors contributed equally to this work: Jun Shu, Yanmin Zhang and Hong Du\\
\textrm{\Letter} e-mail: jshen@bit.edu.cn; 
jinbopeng@sjtu.edu.cn; 
rzhong@sjtu.edu.cn; 
ge.he@bit.edu.cn
}

\begin{abstract}
Electron-phonon coupling is usually understood in terms of electronic quasiparticles interacting with dispersive lattice vibrations. Much less is known about the complementary limit in which the relevant phonon mode is itself localized or weakly dispersive. Here we investigate this regime in the kagome compound \RNS, which undergoes an unconventional insulating transition near $T^{*} \approx$ 260-280~K. Combining polarization-resolved Raman spectroscopy with temperature-dependent x-ray diffraction, scanning tunneling microscopy, and electrical, thermal, and magnetic measurements, we show that the transition involves electronic localization without a conventional structural or magnetic order parameter. Raman spectra reveal a giant Franck--Condon progression associated with a nearly dispersionless 333.7~cm$^{-1}$ phonon, decorated by an equally spaced comb-like fine structure with a characteristic spacing of 40.6~cm$^{-1}$. The comb spacing is insensitive to magnetic field, whereas its spectral weight is strongly field tunable. Rather than treating either hierarchy alone as pure phonon effect, we interpret their nested coexistence as evidence for a strongly coupled electron-vibrational manifold involving a localized lattice coordinate. These results identify dispersionless phonons as an active route to vibronic correlations in solids and suggest that such electron-vibrational self-trapping is closely associated with the insulating phase of \RNS.

\end{abstract}

\maketitle

\section{Introduction}
Electron--phonon coupling originates from the modulation of the crystalline potential by lattice vibrations. This interaction underlies quasiparticle mass enhancement and Kohn anomalies \cite{Giustino:2017,Kohn:1959}, phonon-mediated superconductivity \cite{Bardeen:1957,Allen:1975}, charge-density-wave instabilities \cite{Gruner:1988}, and polaronic quasiparticles \cite{Holstein:1959,Devreese:2009}. Which regime is realized is controlled not only by the microscopic coupling matrix element, but also by the relevant electronic bandwidth and phonon energy scales.

This energy-scale perspective has become especially important in flat- and narrow-band materials. Heavy-fermion compounds provide a canonical example in which electronic quasiparticles acquire extremely small kinetic energy \cite{Kirchner:2020}, while moiré materials realize tunable narrow bands and correlated superconducting states \cite{Bistritzer:2011,Cao:2018,Tilak:2021}. Kagome compounds similarly host narrow or flat electronic bands together with Dirac fermions and van Hove singularities, enabling charge order, superconductivity, and topological responses \cite{Liu:2020,Li:2018,Kang:2020,Yin:2019,HuJY:2023,Checkelsky:2024}. In these systems, lattice fluctuations can have an amplified effect on the electronic state, producing enhanced lattice dressing, polaron formation, and non-perturbative electron--phonon responses \cite{Yin:2020,Korshunov:2023,Chen:2024,Lihm:2025}. Most of this work, however, focuses on the consequences of reduced electronic bandwidth while treating the phonons as conventional dispersive modes.

\begin{figure*}[ht!]
  \centering
  \includegraphics[width=17cm]{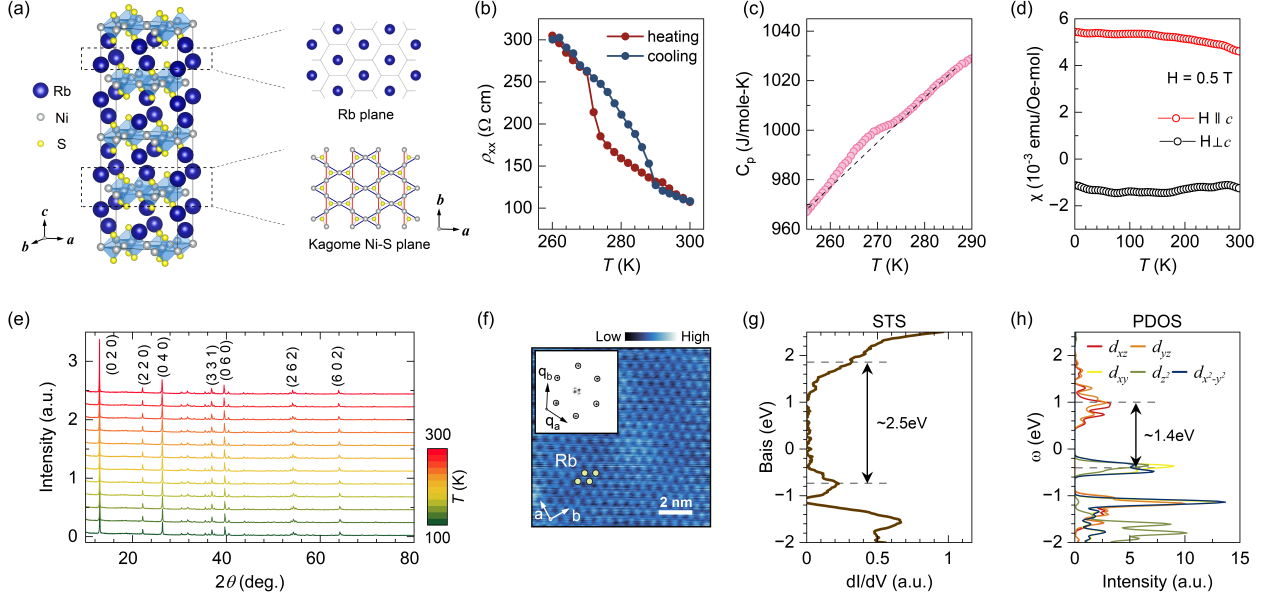}
  \caption{\textbf{Crystal structure and phase transition of \RNS.} 
  (a), Three-dimensional crystal structure (left) highlighting the triangular Rb sublattice and the kagome Ni-S sublattice (right). Rb atoms are shown as blue spheres, while Ni and S atoms are represented by gray and yellow spheres, respectively.
 (b), Temperature dependence of the electrical resistivity. A pronounced anomaly is observed near 280 K, accompanied by a clear thermal hysteresis between cooling and warming processes.
  (c), Temperature-dependent specific heat, showing a weak anomaly with small latent heat across the transition, consistent with its weakly first-order nature. The dashed line is included as a guide to the eye.
  (d), Magnetic susceptibility as a function of temperature, exhibiting no clear signature of long-range magnetic ordering.
  (e), Temperature-dependent X-ray diffraction curves in power samples from 100\,K to 300\,K, revealing no detectable structure change across the transition.
  (f), Atomically resolved STM topography of the cleaved Rb-terminated surface of \RNS. The yellow markers indicate the positions of the surface Rb atoms. Inset: Corresponding fast Fourier transform (FFT) of the STM image. The Bragg peaks are highlighted by black circles. 
  (g)~Scanning tunneling spectroscopy (STS) measured in a defect-free region. The double-headed arrow denotes the experimentally determined band gap. 
  (h)~Orbital-resolved density of states from DFT calculations. The double-headed arrow indicates the calculated band-gap size. 
  }
  \label{fig:phasetransition}
\end{figure*}

A complementary regime emerges when the relevant lattice excitation is itself localized or weakly dispersive. Theoretically, local electron-phonon models such as the Holstein model have long shown that coupling to dispersionless Einstein phonons can produce polaron formation and ordered electronic states \cite{Holstein:1959}. Experimentally, flat or nearly flat phonon modes have begun to emerge as active participants in electron-phonon physics. In CoSn, scanning tunneling spectroscopy (STS) revealed strong coupling between electronic quasiparticles and a kagome flat-band phonon \cite{Yin:2020}; in ScV$_6$Sn$_6$, inelastic x-ray scattering and theory showed that electron--phonon coupling drives the softening of a rather flat phonon mode associated with unconventional charge order \cite{Korshunov:2023,Hu:2025}. These examples suggest that a flat phonon should not be viewed merely as a high-density-of-states lattice mode, but as a localized vibrational degree of freedom capable of reorganizing the coupled electronic response. This raises the central question: can coupling to a localized or weakly dispersive phonon generate emergent electron-vibrational states beyond conventional quasiparticle dressing?

Here, we address this question in \RNS by combining temperature-dependent x-ray diffraction, scanning tunneling microscopy, electrical, thermal, and magnetic transport, and temperature- and magnetic-field-dependent, polarization-resolved Raman scattering. The structural and transport measurements reveal a weak first-order insulating phase transition. Raman spectroscopy reveals an intense Franck--Condon progression associated with a nearly dispersionless phonon mode at 333.7~cm$^{-1}$, together with equidistant sideband combs spaced by 40.6~cm$^{-1}$ on top of the overtone structure. The fixed comb spacing, its placement within the Franck--Condon hierarchy, and its strong field-dependent spectral weight cannot be accounted for by simple phonon replicas alone. Instead, these observations point to a strongly coupled electron--vibrational manifold in which a localized lattice coordinate reorganizes electronic excitations and produces an additional low-energy hierarchy in the Raman response. Our results therefore suggest that vibronic self-trapping participates in the electronic localization accompanying the insulating transition in \RNS, without requiring a conventional structural or magnetic order parameter.\\

\begin{figure*}[ht!]
  \centering
  \includegraphics[width=15cm]{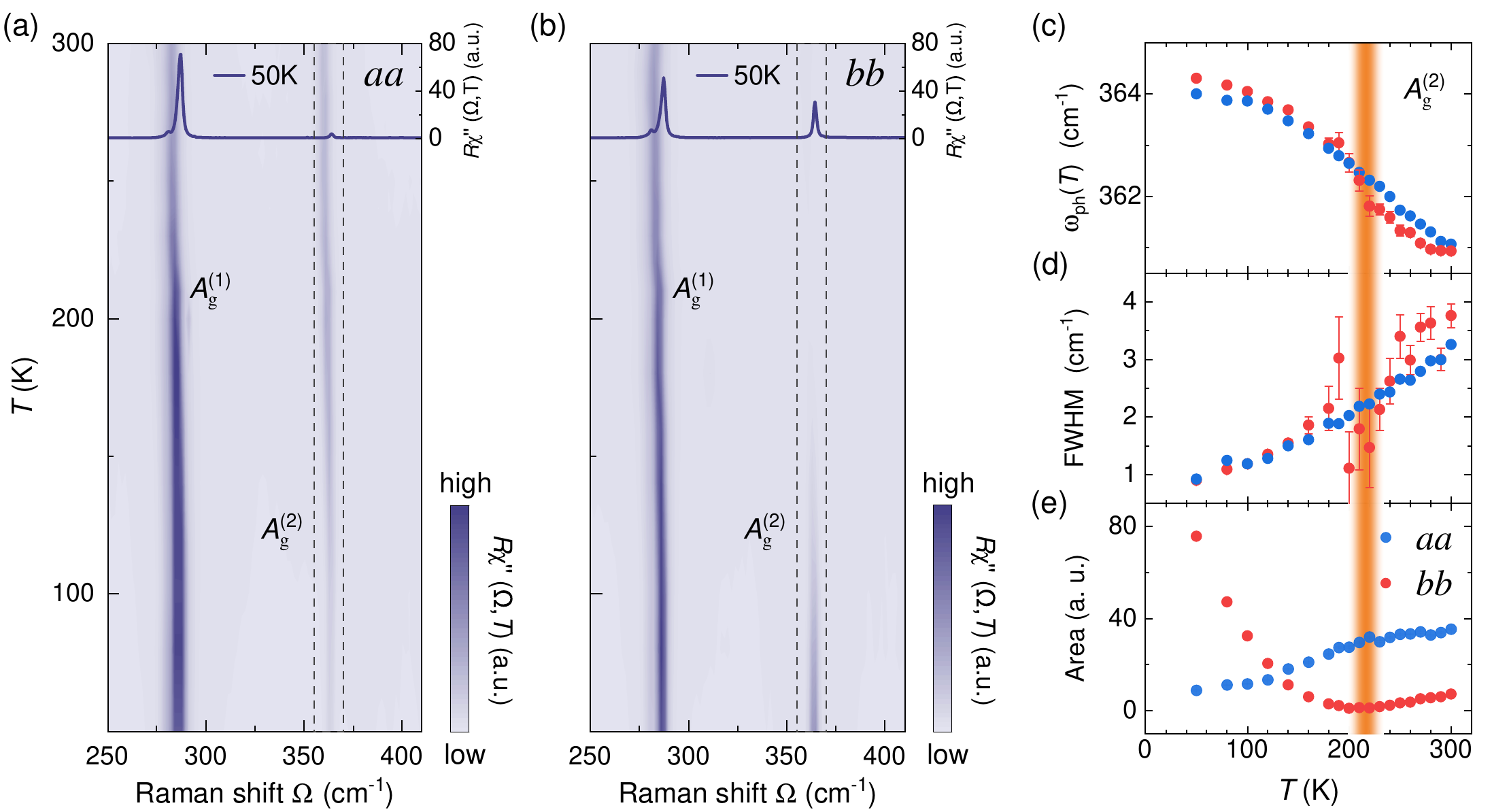}
\caption{\textbf{Temperature-dependent Raman-active phonons.} 
(a) and (b)~Temperature evolution of the Raman spectra in $aa$ and $bb$ polarization configurations, respectively. Insets: Raman response measured at 50\,K in the \textit{aa} and \textit{bb} polarization configuration, respectively.
(c)~Temperature dependence of the \Agz phonon frequency obtained from $aa$ and $bb$ channels.
(d)~Temperature dependence of the \Agz phonon linewidth (full width at half maximum, FWHM).
(e)~Temperature dependence of the integrated Raman intensity (peak area) of the \Agz phonon.
The phase transition is indicated by the yellow region centered around 220~K, which is approximately 50~K lower than the transition temperature inferred from transport measurements, reflecting the different sensitivities of the two probes to a first-order transition.}
\label{fig:phonon}
\end{figure*}


\section{Results and Discussion}

\subsection{1. Weak first-order phase transition}

\RNS has been reported to crystallize in two structural variants, a lower-symmetry orthorhombic phase with space group $Fmmm$ (No.~69) and a higher-symmetry hexagonal phase with space group $P6_3/mmc$ (No.~194). X-ray diffraction refinement indicates that our sample adopts the orthorhombic structure, featuring a slightly distorted kagome lattice (see Supplementary Material A for details), similar to that reported for Cs$_2$Ni$_3$S$_4$ \cite{Villalpando:2024}. The crystal structure consists of quasi-two-dimensional Ni-S kagome layers, with adjacent layers separated by two intervening Rb atomic planes [Fig.~\ref{fig:phasetransition}(a)]. This layered architecture effectively decouples the Ni--S planes, preserving the essential electronic characteristics of the kagome lattice.



\RNS undergoes an insulating transition at $T^* \approx 260$--$280$\,K \cite{DuH:2024}, but its microscopic origin remains unresolved \cite{Villalpando:2024}. Temperature-dependent resistivity, $\rho(T)$, shows a pronounced anomaly at $T^*$ with clear thermal hysteresis [Fig.~\ref{fig:phasetransition}(b)], while the specific heat exhibits only a weak anomaly with a small latent heat across the transition temperature [Fig.~\ref{fig:phasetransition}(c)], indicating a weakly first-order transition. Despite the large transport anomaly, temperature-dependent x-ray diffraction detects no structural phase transition across $T^*$ [Fig.~\ref{fig:phasetransition}(e)]. Scanning tunneling microscopy (STM) further supports the absence of lattice reconstruction down to 77~K. The real-space topography preserves the high-temperature lattice periodicity, and the corresponding FFT maps show only the fundamental Bragg peaks, with no additional peaks associated with superlattice modulation or charge ordering [Fig.~\ref{fig:phasetransition}(f)]. STS confirms the insulating ground state with a gap of $\sim$2.5~eV [Fig.~\ref{fig:phasetransition}(g)], and the valence-band is dispersionless as confirmed by previous ARPES measurements \cite{Nawai:2004}. Density-functional theory (DFT) calculations also yield an insulating state, albeit with a smaller gap of approximately 1.4~eV [Fig.~\ref{fig:phasetransition}(h)]. The unoccupied states are dominated by Ni $d_{xz}$ and $d_{yz}$ orbitals, whereas the occupied manifold mainly consists of Ni $d_{xy}$, $d_{x^2-y^2}$, and $d_{z^2}$ orbitals, consistent with previous reports \cite{Bahadur:2024}.

These results strongly constrain the origin of the transition. The absence of detectable lattice reconstruction rules out a conventional structural transition or charge-density-wave instability as the primary mechanism. Magnetism is also unlikely to be the dominant driving force: although the magnetic ground state of \RNS remains debated, with early reports suggesting weak ferromagnetism \cite{Kato:1998} and later studies favoring a paramagnetic ground state \cite{Fukamachi:1999}, our susceptibility measurements show no clear anomaly near $T^*$ [Fig.~\ref{fig:phasetransition}(d)], and the reported effective moment per Ni atom is extremely small \cite{DuH:2024}. Taken together, the transport, thermodynamic, structural, spectroscopic, and magnetic data point to an unconventional route to electronic localization, motivating a closer examination of electron--lattice coupling in this system.

\subsection{2. Phonon anomaly}
To further examine whether the observed anomalies are accompanied by changes in lattice symmetry, we turn to polarization-resolved Raman spectroscopy in \RNS. Factor-group analysis predicts three $A_g$ and one $B_{1g}$ Raman-active phonon modes for $ab$-plane light polarization geometries (see Supplementary Material~B for details). Experimentally, we resolve two $A_g$ modes and one $B_{1g}$ mode, labeled \Agl, \Agz, and \Blg, respectively [Insets of Fig.~\ref{fig:phonon}(a) and (b)]. These modes are observed near 286~cm$^{-1}$ (\Agl), 362~cm$^{-1}$ (\Agz), and 280~cm$^{-1}$ (\Blg), consistent with pervious Raman results \cite{Hasegawa:2004}. Among the three prominent Raman-active phonons, the \Agz mode exhibits pronounced anisotropy between the $a$- and $b$-axis polarization configurations (results of other phonons are summarized in Supplementary Material~C). Upon cooling, the intensity of the \Agz phonon shows markedly different behavior in the two channels. It is strongly enhanced below $\sim 200$~K in the $bb$ configuration, while it decreases monotonically in the $aa$ configuration, as shown in Fig.~\ref{fig:phonon}\,(b) and (a). Quantitatively, fitting the phonon line shape with a Voigt profile (a Lorentzian convoluted with a Gaussian) allows us to extract the phonon frequency, full width at half maximum (FWHM), and integrated intensity. Across the transition, the \Agz mode exhibits a clear frequency hardening of approximately $\Delta\omega \approx 1$~cm$^{-1}$ in the $bb$ channel [Fig.~\ref{fig:phonon}\,(c)], accompanied by a pronounced linewidth fluctuation of about 2~cm$^{-1}$ [Fig.~\ref{fig:phonon}\,(d)]. The integrated intensity of the \Agz phonon displays a nonmonotonic temperature dependence, decreasing upon cooling before increasing again near 200\,K [Fig.~\ref{fig:phonon}\,(e)]. In contrast, no transition-related anomalies are observed in the $aa$ channel in terms of mode frequency, line width and the intensity. We note that the characteristic phase transition temperature $T^*$ inferred from Raman measurements is approximately 50~K lower than that determined from transport experiments. This offset likely reflects the different sensitivities of the two probes to the phase transition, particularly in the presence of strong hysteresis and phase coexistence expected for a first-order transition. \\


\begin{figure*}[ht!]
  \centering
  \includegraphics[width=15cm]{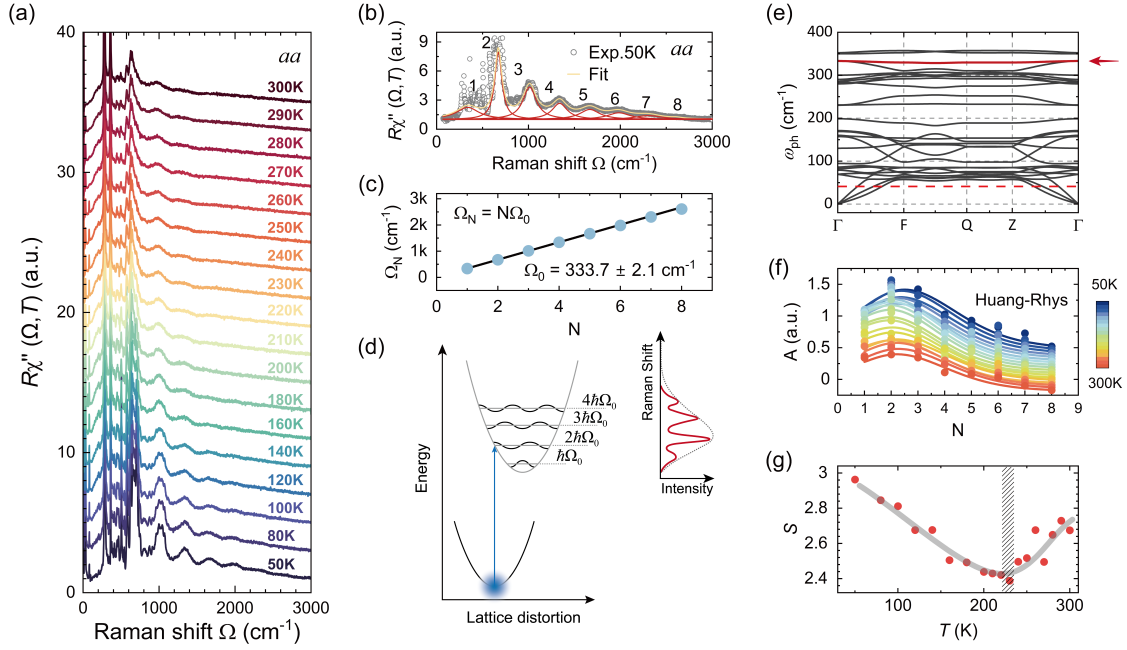}
\caption{\textbf{Phonon overtones in \RNS.} 
(a)~Raman spectra at various temperatures, highlighting phonon overtones.
(b)~Representative fit of the overtone profile up to the eighth order.
(c)~Extracted overtone peak positions as a function of overtone order. The peak positions follow a linear dependence with a slope of $333.7\pm2.1~\rm{cm}^{-1}$, corresponding to the base phonon frequency.
(d)~Schematic illustration of the Franck-Condon progression. In the excited electronic state, the ionic potential (gray parabola) is displaced relative to the ground state potential (black parabola) due to electron-phonon coupling, giving rise to multiphonon excitations. The resulting Poisson distribution of Raman intensities is shown on the right.
(e)~Calculated phonon dispersion from density functional theory. The $333.5~\rm{cm}^{-1}$ dispersion-less phonon is indicated in red arrow. The red dashed line indicates the energy at 40~\wn, corresponding to the characteristic energy of the sidebands.
(f)~Overtone peak intensities as a function of order number at different temperatures. Solid lines are fits using the Franck-Condon model. For clarity, data other than those measured at the
highest temperature are shifted vertically.
(g)~Temperature dependence of the Huang-Rhys factor extracted from Franck-Condon fits. The shaded region indicates the phase transition.}
\label{fig:overtone}
\end{figure*}


\subsection{3. Franck-Condon progression}

Extended Raman spectra reveal a series of gradually damped peaks accompanied by a broad continuum, which become strongly enhanced below $T^*$. These features are most prominent in the $aa$ and $bb$ polarization configurations, with barely detectable signals in the $ab$ channel (Supplementary Material D1). Figure~\ref{fig:overtone}\,(a) shows the temperature evolution of these modes. Remarkably, we observe a well-defined sequence of damped peaks extending up to the eighth order, with peak positions located at 337.9~cm$^{-1}$ (1st order), 671.4~cm$^{-1}$ (2nd), 1011.9~cm$^{-1}$ (3rd), 1338.6~cm$^{-1}$ (4th), 1667.3~cm$^{-1}$ (5th), 1983.0~cm$^{-1}$ (6th), 2306.1~cm$^{-1}$ (7th), and 2608.9~cm$^{-1}$ (8th), as summarized in Fig.~\ref{fig:overtone}\,(b). These peak positions follow an essentially perfect linear dependence on the order, yielding a slope of $333.7\pm2.1~\rm{cm}^{-1}$, which identifies the underlying base mode frequency [Fig.~\ref{fig:overtone}\,(c)] (see Supplementary Material D2 and D3 for details).

Luminescence can be ruled out as the origin of these features, as these peaks are reproducibly observed using multiple excitation wavelengths, including blue (473~nm), green (532~nm), and red (633~nm) lasers (Supplementary Material E). The remaining natural interpretation is multiphonon scattering (phonon overtone), a phenomenon observed in many systems, such as monochalcogenides (e.g., CdS \cite{Venkateswaran:1985} and ZnTe \cite{Feng:1991}), perovskites (e.g., Na-doped Cs$_2$AgInCl$_6$ \cite{XuKX:2022}, Cs$_2$NaFeCl$_6$ \cite{Kelley:2013} and (PAE)$_2$PbI$_4$ \cite{Dyksik:2024}), semiconducting magnets (e.g., CrI$_3$ \cite{Jin:2020}), and most recently, spinel oxide superconductor LiTi$_2$O$_4$ \cite{Hasan:2026} and Hg-based cuprate superconductors \cite{Hong:2025}.

Several distinct mechanisms can, in principle, give rise to multiphonon Raman features and must be carefully distinguished. In the harmonic approximation, multiphonon scattering arises from independent phonon creation processes subject to momentum conservation, typically producing complex peak structures with rapidly diminishing intensity at higher orders \cite{Carvalho:2017, Jin:2020}. In the cascade model, multiphonon peaks originate from successive energy quanta of a single phonon mode, often yielding a sharp first-order peak followed by an exponential decay in higher orders \cite{Varma:1971}. By contrast, the Huang-Rhys model describes a Franck-Condon process in the strong electron-phonon coupling limit, in which electronic excitation displaces the ionic potential, leading to a Poisson distribution of multiphonon intensities with a characteristic maximum near $n_{\mathrm{max}} \approx S$, where $S$ is the Huang-Rhys factor \cite{Huang:1950, Hopfield:1959} [Fig.~\ref{fig:overtone}\,(d)].

First-principles phonon calculations identify two $\Gamma$-point optical phonon modes near 334~cm$^{-1}$ with $B_{2u}$ and $B_{3u}$ symmetry as the relevant vibrational modes [Fig.~\ref{fig:overtone}\,(e)] (also see Supplementary Material F for details). Notably, the $B_{3u}$-derived mode is nearly dispersionless, potentially forming a local vibration mode. Particularly, this mode originates from the Kagome lattice itself (see Supplementary Material~G for details). The absence of a dominant sharp first-order peak, together with the unusually slow decay of overtone intensities extending up to the eighth order, argues against both the harmonic approximation and cascade scenarios. Furthermore, the base frequency phonon involved is Raman inactive (odd parity), closely analogous to the situation in CrI$_3$, where multiphonon Raman features arise from polaronic excitations \cite{Jin:2020}.

These observations strongly support the Huang--Rhys description of the multiphonon response, placing \RNS in the strong electron-phonon coupling regime. We therefore fit the overtone intensities at various temperatures [Fig.~\ref{fig:overtone}\,(f)] (for fitting details, see Supplementary Material D3) using a Poisson distribution \cite{Langreth:1970},
\begin{equation}
A_n = A_0 \frac{e^{-S} S^n}{n!},
\end{equation}
where $A_n$ is the intensity of the $n$th-order overtone, $A_0$ is a normalization constant, and $S$ is the Huang-Rhys factor. The fits capture the experimental data remarkably well and yield the temperature dependence of $S$ shown in Fig.~\ref{fig:overtone}\,(g). Upon warming, $S$ initially decreases and then slightly increases above $T^*$. At 50~K, we obtain $S \approx 2.9$, approaching the value reported for CrI$_3$ ($S \approx 3.5$) \cite{Jin:2020}. Such a large Huang–Rhys factor points to unusually strong electron–phonon coupling in the present material.\\

\begin{figure*}[ht!]
  \centering
  \includegraphics[width=15cm]{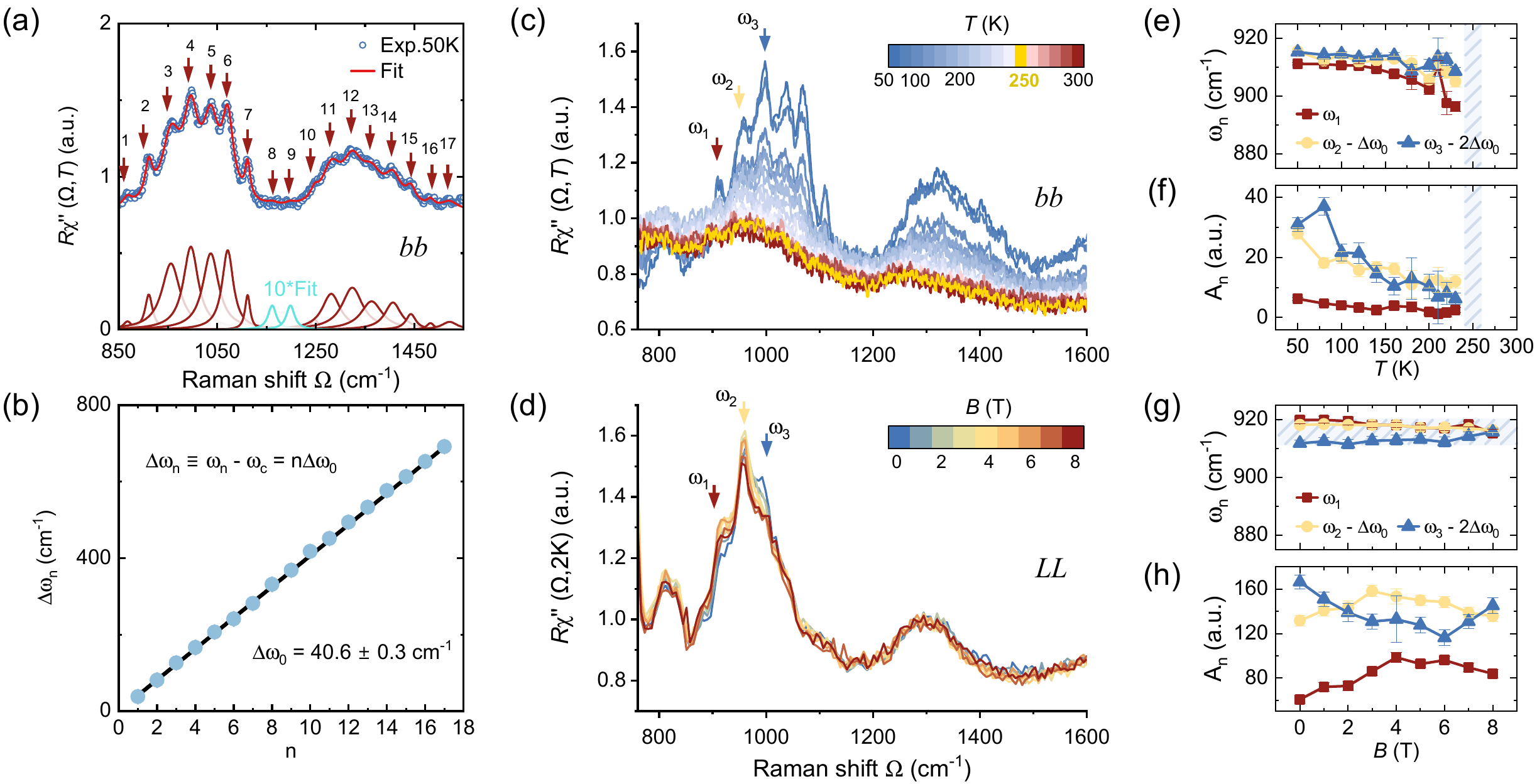}
\caption{\textbf{Sideband structure in \RNS.} 
(a)~Raman spectrum in the $bb$ polarization configuration spanning the third- to fourth-order phonon overtones. The red arrows highlight additional fine-structure features superimposed on the overtone background. Isolated fine-structure peaks obtained after subtracting the fitted phonon overtones. Red curves indicate fits to the individual fine-structure modes.
(b)~Extracted fine-structure peak positions as a function of sideband order. The relative peak positions $\Delta\omega_n \equiv \omega_n - \omega_c$ follow a linear dependence with a slope of 40.6~cm$^{-1}$, where $\omega_n$ denotes the absolute peak positions and $\omega_c$ is a reference energy set to 830~cm$^{-1}$.
(c)~Temperature evolution of the Raman spectra in the third- to fourth-order overtone range, showing the development of fine structures in the $bb$ channel using 532~nm excitation. The phase transition temperature is highlighted in yellow.
(d)~Magnetic-field dependence of the Raman spectra in the same energy range, measured in the $LL$ polarization configuration using 473~nm excitation, with magnetic fields up to 8~T.
(e,f)~Temperature dependence of the fine-structure peak positions $\omega_n$ and integrated peak areas $A_n$, respectively.
(g,h)~Magnetic-field dependence of $\omega_n$ and $A_n$, respectively. The analyzed peaks are indicated with arrows in panels (c) and (d).}
\label{fig:finestructure}
\end{figure*}


\subsection{4. Fine-structure sidebands}

Superimposed on the high-order phonon overtones, we resolve an additional set of well-defined fine-structure peaks, most clearly observed in the parallel polarization configuration ($aa$ and $bb$) within the energy window spanning the third- to fourth-order overtones [Fig.~\ref{fig:finestructure}\,(a)]. The fine structures are much weaker in the $ab$ polarization channels (Supplementary Material~D1). Notably, the fine-structure intensity decreases markedly in the fourth-order overtone and becomes undetectable in the higher overtones. By fitting the features with voigt function we found that these features emerge as a series of narrow ($\bar{\Gamma}\sim20$\,\wn), nearly equidistant peaks. The extracted peak positions exhibit a strictly linear dependence on the sideband index, yielding a uniform energy spacing of $40.6\pm0.3~\mathrm{cm}^{-1}$ [Fig.~\ref{fig:finestructure}\,(b)]. 

The temperature evolution of the fine structures further supports their unconventional nature (Supplementary Material~H1). Upon warming, both the peak energies and integrated intensities gradually decrease [Fig.~\ref{fig:finestructure}\,(e) and (f)], and the fine structures become progressively less resolved [Fig.~\ref{fig:finestructure}\,(c)]. Above the phase transition temperature $T^*$, the sidebands are no longer discernible within experimental resolution, indicating that their existence is intimately tied to the low-temperature phase. 

The comb-like fine structure cannot be understood as a conventional multiphonon process, as there is no phonon mode with a high density of states near 40~cm$^{-1}$ (see Fig.~\ref{fig:overtone}\,(e)). It is also unlikely to originate from folded phonons associated with a large-period superlattice~\cite{Colvard:1980} or from moir\'e systems~\cite{Lin:2018}, since no evidence for such structural modulations is observed in our STM and X-ray measurements.

To further distinguish whether the sidebands stem from pure phononic excitation or mixed with electronic states, we performed field-dependent measurement. To exclude artifacts arising from Faraday rotation or polarization mixing in magnetic field \cite{Jin:2020}, the measurements were performed using circularly polarized light in the $LL$ configuration. As shown in Fig.~\ref{fig:finestructure}\,(d) and (g), the energies of the fine-structure peaks remain essentially unchanged up to 8~T, demonstrating that the characteristic excitation energy scale is robust against Zeeman splitting. In contrast, the peak intensities exhibit a clear and systematic magnetic-field dependence [Fig.~\ref{fig:finestructure}\,(h)]. Specifically, the intensity modulation displays an apparent periodicity of three: the central peak is nearly unchanged (marked as $\omega_2$ in Fig.~\ref{fig:finestructure}\,(d)), while the lower-energy sideband gains intensity ($\omega_1$) and the higher-energy sideband loses spectral weight ($\omega_3$) with increasing field. Complementary measurements in cross-circular polarization ($RL$) further confirm the intrinsic nature of the effect, where the intensity against field relation reverses (Supplementary Material~H2 and H3). This selective redistribution of spectral weight, without a corresponding energy shift, points to field-tunable quantum interference among symmetry-related states rather than conventional magnetic excitations.\\

\subsection{5. Discussion}

The central result is the simultaneous appearance of two spectral hierarchies associated with the nearly dispersionless B$_{3u}$ phonon. The Franck-Condon progression, characterized by a Huang-Rhys factor $S\approx3$, establishes a large lattice displacement coupled to the electronic excitation. Superimposed on each overtone is an equally spaced fine structure with a characteristic spacing of 40.6~cm$^{-1}$. Either feature can occur separately. Franck-Condon progressions reflect strong electronic--vibrational coupling, as demonstrated in CrI$_3$ \cite{Jin:2020}, whereas narrow frequency combs can arise from long-lived nonlinear phonons, as recently observed in CrGeTe$_3$ and CrSiTe$_3$ \cite{Chen:2025}. In Rb$_2$Ni$_3$S$_4$, however, the comb is embedded within the Franck--Condon manifold. This nested structure indicates that strong lattice dressing and phase-correlated vibrational dynamics belong to the same excitation.

This response is distinct from several established limits of electron--phonon physics. In a conventional Franck--Condon process, the optical excitation connects two displaced potential-energy surfaces and produces a vibrational progression whose intensity distribution reflects the lattice reorganization energy; such a progression alone does not require a long-lived internal phase structure \cite{Huang:1950,Hong:2025,Jin:2020}. In a purely phononic frequency comb, the equally spaced peaks can arise from nonlinear dynamics of a localized phonon mode with restricted decay channels, without requiring strong electronic participation \cite{Fermi:1955,Berman:2005,Chen:2025}. Rb$_2$Ni$_3$S$_4$ combines elements of both responses in a single Raman spectrum. The Franck--Condon envelope indicates that the excitation is electronically active and strongly lattice dressed, while the embedded comb and its magnetic-field-dependent spectral weight indicate that the additional energy hierarchy is coupled to electronic degrees of freedom rather than being a purely phononic replica series. We therefore use the term vibronic state in an operational sense, a strongly coupled electron-vibrational excitation whose Raman response cannot be decomposed into an electronic quasiparticle plus independent phonon replicas. This interpretation provides a constrained phenomenology for the coupled electronic and local vibrational degrees of freedom in \RNS \cite{Alexandrov:2010} (see Supplementary Material I for details).

The nearly flat phonon is the key ingredient that makes this regime possible. Because the mode is weakly dispersive, it behaves as a localized vibrational coordinate with reduced propagation-induced dephasing \cite{Barman:2004, Chen:2025}. This favors repeated coupling to the same electronic degree of freedom and restricts the decay phase space of the vibrational motion. In Rb$_2$Ni$_3$S$_4$, the strong coupling between electrons and the dispersionless B$_{3u}$ produces both the large lattice reorganization encoded by the Huang--Rhys factor and the long-lived internal spectral structure encoded by the comb. The magnetic-field dependence reinforces this interpretation. The comb spacing remains essentially unchanged, indicating that its internal energy scale is set by the lattice mode, whereas the spectral weight changes strongly with field, revealing sensitivity to the electronic or magnetic sector. This separation is difficult to reconcile with a purely phononic comb, but follows naturally for a vibronic excitation whose energy hierarchy is phonon-defined while its Raman matrix element contains electronic character.

The same electron-vibrational coupling offers a plausible route to the insulating transition. Conventional insulating transitions are often framed in terms of electron correlation, as in a Mott transition \cite{Imada:1998}, or Fermi-surface instability coupled to lattice distortion, as in a Peierls or charge-density-wave transition \cite{Gruner:1988}. Strong electron-phonon coupling can also produce polaronic localization, in which the lattice distortion follows the carrier, renormalizes its motion, and eventually traps it \cite{Holstein:1959,Jin:2020,Dyksik:2024,Hasan:2026}. In \RNS, the observed Franck--Condon displacement indicates that the electronic excitation strongly reorganizes a nearly flat local phonon coordinate, while the embedded comb suggests that this coordinate retains structured dynamics rather than acting as a simple incoherent dressing cloud. Taken together, these observations are consistent with vibronic self-trapping contributing to the insulating phase. The coupling to the nearly dispersionless phonon may suppress electronic hopping and promote carrier localization through an emergent electron-vibrational state resolved directly in the Raman spectrum.

\section{Conclusion}
To summarize, Rb$_2$Ni$_3$S$_4$ exhibits a giant Franck--Condon progression decorated by an equally spaced comb-like fine structure. Their coexistence identifies a regime beyond conventional Franck--Condon scattering, ordinary polaron formation, and purely phononic comb dynamics. A nearly dispersionless phonon supplies a localized vibrational coordinate that strongly reorganizes the electronic excitation, while the magnetic-field dependence of the comb spectral weight shows that the Raman response retains electronic character. Taken together with the structural, microscopic, and transport signatures across the transition, these results support a picture in which a coupled electron--vibrational manifold is closely linked to electronic localization in the insulating phase. More broadly, they establish flat phonons as a route to collective vibronic correlations in solids.\\

\section{Acknowledgments}
We thank Rudi Hackl, Chandra Verma, Ruizhen Huang, Yuanji Xu and Zengyi Du for fruitful discussions. This work is supported by the National Natural Science Foundation of China (Grants No. 12474473, 12374191, 12374187, 12104490) and the National Key Basic Research Program of China (Grants No. 2024YFF0727103). J.P. acknowledges support from the Ministry of Science and Technology of China (Grant No. 2023ZD0301300) and TDLI Start-up Grant. L.Q. acknowledges the support by the Youth Beijing Scholars program (Grant No. 93) and the Project of Construction and Support for high-level Innovative Teams of Beijing Municipal Institutions (BPHR20220124). J.Y. and K.J. acknowledge the support by CAS Project for Young Scientists in Basic Research (Grant No. 2022YSBR-048). This work is supported by the Center for Materials Genome and the Synergetic Extreme Condition User Facility (SECUF, https://cstr.cn/31123.02.SECUF). \\

\section{Data  Availability}
All relevant data that support the findings of this study are presented in the manuscript and supplementary information file. All data are available upon request from the corresponding authors.

\section{Appendix A: Sample synthesis} 
High-quality \RNS single crystals were synthesized by fusion reaction under a flowing Ar atmosphere. Mixtures of pre-dried Rb$_{2}$CO$_{3}$(99.994\%), Ni (99.9999\%), and S ($>$99.99\%) in a molar of 1: 5: 12 were placed into alumina crucibles as starting materials. The crucibles were then heated at a rate of 9 °C min$^{-1}$ to 1000 °C and held at this temperature for 0.5 $\sim$ 3 h, and then furnace-cooled to room temperature. Single crystals with layered morphology can easily be obtained by mechanical exfoliation from the surface of the reaction products. \\

\section{Appendix B: Light scattering}

Inelastic light-scattering measurements were performed using two commercial single-stage Raman spectrometers (Horiba HR-800). One system is integrated with a closed-cycle, liquid-helium-free optical cryostat, enabling temperature control from 50 to 350~K. The other is coupled to an attocube cryogenic dewar, allowing measurements down to 2~K under magnetic fields up to 9~T. Samples were freshly cleaved using the Scotch-tape method prior to mounting on the cryostat cold finger. Excitation was provided by solid-state lasers with wavelengths of 473, 532, and 633~nm. The incident laser power was typically set to $P = 3.0$~mW, corresponding to an estimated local heating of approximately 1--2~K per milliwatt. Integration times ranged from 480 to 600~s for each spectrum. Polarization configurations are denoted as $aa$, $ab$, and $bb$, where $a$ and $b$ correspond to the crystallographic axes of the sample. The Raman response is presented in terms of the Raman susceptibility, $R\chi^{\prime\prime}(\Omega,T) = \pi \left[1 + n(\Omega,T)\right]^{-1} S(q \approx 0, \Omega)$, where $R$ is an experimental constant, $\chi^{\prime\prime}(\Omega,T)$ is the imaginary part of the Raman response function, $S(q \approx 0, \Omega)$ is the dynamical structure factor proportional to the photon scattering rate, and $n(\Omega,T)$ is the Bose-Einstein distribution function \cite{Devereaux:2007}.
\\

\section{Appendix C: STM/STS} Scanning tunneling microscopy (STM) experiments were conducted at both 300~K and 77~K using a Createc (Germany) STM system. A fresh tungsten tip was fabricated by electrochemical etching (0.3~$\mu$m diameter wire) and calibrated on Au(111) prior to each measurement. Single crystals of Rb$_2$Ni$_3$S$_4$ were cleaved in situ to expose fresh surfaces for imaging. Two cleaving methods were used: (1) a post-top cleaving method performed within the main STM chamber (base pressure $< 2 \times 10^{-10}$~mbar) at 77~K, and (2) cleavage using Kapton tape in the load-lock chamber (base pressure $< 1 \times 10^{-7}$~mbar) at room temperature. No significant difference was observed between the surfaces prepared by these two methods. The $\mathrm{d}I/\mathrm{d}V$ spectra were obtained through a lock-in detection of the tunnelling current with a modulation voltage amplitude of 8~mV or 5~mV and a modulation frequency of 914.7~Hz. All voltages refer to the sample bias with respect to the tip.\\

\section{Appendix D: DFT simulations}
Our study is based on density functional theory (DFT) calculations performed using the Vienna \textit{ab initio} Simulation Package (VASP)~\cite{vasp1,vasp2}. 
The generalized gradient approximation (GGA) with the Perdew--Burke--Ernzerhof (PBE) functional is employed for the exchange-correlation term~\cite{pbe}. 
To describe the core electron interaction, the projector augmented wave (PAW)~\cite{paw1,paw2} are used. Brillouin zone integrations are performed using a Monkhorst-Pack \textit{k}-point grid of $5 \times 5 \times 5$~\cite{monkhorst}. 
The energy cutoff is set to be 520~eV. 
The total energy is converged to better than $10^{-5}$~eV, and the geometries are fully relaxed until the force on each atom is less than 0.01~eV/\AA. The calculated lattice constants of \(\mathrm{Rb_2Ni_3S_4}\) 
(\(a = 8.486\)~\AA, \(b = 7.478\)~\AA, and \(c = 5.769\)~\AA) 
are used to investigate its electronic band, density of states, and phonon band dispersions. The phonon band dispersions are calculated in a $2 \times 2 \times 2$ supercell contains 16 Rb atoms, 24 Ni atoms and 32 S atoms using the PHONOPY package~\cite{phonopy,phonopy2}. \\

\bibliography{Rb2Ni3S4_references_20260123,SI}

\end{document}